\documentclass{article}
\usepackage[utf8]{inputenc}
\usepackage{amsmath}
\usepackage{amssymb}
\usepackage{tensor}
\usepackage{mathrsfs}
\usepackage{cite}
\usepackage{multirow}
\usepackage{framed}
\usepackage{graphicx}
\usepackage[colorlinks=true,urlcolor=blue,anchorcolor=blue,citecolor=blue,filecolor=blue,linkcolor=blue,menucolor=blue,linktocpage=true,pdfproducer=medialab,pdfa=true]{hyperref}
\usepackage{calc}
\usepackage[symbol]{footmisc}

\newcommand{\Op}{\mathcal{O}}

\begin{document}

\numberwithin{equation}{section}

\begin{titlepage}
\noindent
\hfill August 2025\\
\vspace{0.6cm}
\begin{center}
{\LARGE \bf 
    Alien operators for PDF evolution\footnote{Presented at the European Physical Society Conference on High Energy Physics (EPS-HEP2025), Marseille, France, 7-11 July 2025.}}\\ 
\vspace{1.4cm}

\large
S. Van Thurenhout$^{\, a}$\footnote{Speaker},
G. Falcioni$^{\, b,c}$,
F. Herzog$^{\, d}$ and
S. Moch$^{\, e}$
\\
\vspace{1.4cm}
\normalsize
{\it $^a$ HUN-REN Wigner Research Centre for Physics, Konkoly-Thege Mikl\'os u. 29-33, 1121
Budapest, Hungary}\\
\vspace{4mm}
{\it $^b$ Dipartimento di Fisica, Universit\`a di Torino, Via Pietro Giuria 1, 10125 Torino, Italy}\\
\vspace{4mm}
{\it $^c$ Physik-Institut, Universit\"at Z\"urich, Winterthurerstrasse 190, 8057 Z\"urich, Switzerland}\\
\vspace{4mm}
{\it $^d$ Higgs Centre for Theoretical Physics, School of Physics and Astronomy, \\ \vspace{0.5mm}
The University of Edinburgh, Edinburgh EH9 3FD, Scotland, UK}\\
\vspace{4mm}
{\it $^e$II.~Institute for Theoretical Physics, Hamburg University\\
\vspace{0.5mm}
Luruper Chaussee 149, D-22761 Hamburg, Germany}\\
\vspace{1.4cm}

{\large \bf Abstract}
\vspace{-0.2cm}
\end{center}
Understanding the scale dependence of parton distribution functions is vital for precision physics at hadron colliders. The well-known DGLAP evolution equation relates this scale dependence to the QCD splitting functions, which can be calculated perturbatively in terms of the anomalous dimensions of leading-twist gauge-invariant operators. The computation of the latter in general requires one to take into account contributions of gauge-variant (or alien) operators. In this talk, we discuss the systematic study of these alien operators at arbitrary spin. Specifically, using generalized BRST symmetry relations, we derive the one-loop couplings and Feynman rules of the aliens necessary to perform the operator renormalization up to four loops in QCD. This provides an important step towards the determination of the four-loop splitting functions which will be of significant phenomenological importance at future colliders.
\vspace*{0.3cm}
\end{titlepage}

\section{Introduction}
\label{sec:intro}
The splitting functions of quantum chromodynamics (QCD) play an important r\^ole for high-precision phenomenology at hadron colliders, as they determine the scale dependence of the parton distribution functions (PDFs). Their Mellin transforms, which correspond to the anomalous dimensions of the leading-twist gauge-invariant operators that define the distributions, can be determined from the ultraviolet singularities of partonic off-shell operator matrix elements (OMEs). The renormalization of these OMEs requires one to take into account unphysical contributions coming from so-called \textit{alien} operators. These aliens have couplings which are constraint by fundamental symmetries. In this talk, we review how these constraints can be used to reconstruct the full $N$-dependence of the couplings and, consequently, the all-$N$ alien Feynman rules.

\section{Splitting functions and aliens}
\label{sec:Split}
The scale dependence of the PDFs is described by the well-known DGLAP equation \cite{Gribov:1972ri,Altarelli:1977zs,Dokshitzer:1977sg},
\begin{equation}
    \frac{\text{d} f_i(x,\mu^2)}{\text{d} \ln{\mu^2}} = \int_x^1 \frac{\text{d}y}{y} P_{ij}(y)f_j\left(\frac{x}{y},\mu^2\right)\,.
\end{equation}
Contrary to the PDFs themselves, the splitting functions $P_{ij}(y)$ are perturbative quantities. Here we will focus on their Mellin transforms, i.e., on the operator anomalous dimensions,
\begin{equation}
\gamma_{ij} = -\int_0^1 \text{d} x\, x^{N-1}P_{ij}(x)\,,
\end{equation}
which admit a perturbative expansion in terms of the reduced strong coupling constant $a_s=\alpha_s/(4\pi)$,
\begin{equation}
\gamma_{ij} = a_s\,\gamma_{ij}^{(0)} +a_s^2 \gamma_{ij}^{(1)} + ...\,.
\end{equation}
In practice, they can be extracted from the renormalization of the \textit{partonic} matrix elements of the leading-twist operators in off-shell kinematics. When doing so, it is well-known that mixing with non-gauge-invariant (alien) operators needs to be taken into account \cite{Dixon:1974ss,Kluberg-Stern:1974nmx,Kluberg-Stern:1975ebk,Joglekar:1975nu,Joglekar:1976eb,Joglekar:1976pe}. The latter receive contributions from ghost fields on the one hand and field equations of motion on the other. Recently, a method to consistently reconstruct the functional form of the aliens to any loop-order was derived in \cite{Falcioni:2022fdm}. In that approach, the aliens are derived using a generalized gauge symmetry of the QCD Lagrangian, which is then promoted to a generalized (anti-)BRST symmetry. Each alien operator features a coupling that corresponds to the renormalization constant that characterizes the mixing of the physical operators into the alien. It turns out that these couplings obey certain constraints which are $N$-dependent. In \cite{Falcioni:2022fdm,Falcioni:2024xyt}, the latter were solved for \textit{fixed} moments $N\leq 20$, leading to expressions for \textit{all} four-loop splitting functions up to $N=20$ \cite{Falcioni:2023luc,Falcioni:2023vqq,Falcioni:2023tzp,Falcioni:2024xyt,Falcioni:2024qpd} \footnote[2]{The $n_f^2$ contributions to the four-loop pure-singlet splitting functions were obtained in \cite{Gehrmann:2023cqm} using a different framework, which directly determines the alien counterterms \cite{Gehrmann:2023ksf}.}. The solution of the constraints for \textit{arbitrary} values of $N$ was presented in \cite{Falcioni:2024xav}, and we briefly summarize the findings of that work below.

\section{Solving the all-$N$ constraints}
\label{sec:allN}
To illustrate the functional form of the constraints on the alien couplings, and their power, we consider the following ghost operators,
\begin{align}
    \Op_{c}^{(N),I} =\,&\, -\eta(N) (\partial\overline{c}^{a})(\partial^{N-1}c^{a})\,,\\
    \Op_{c}^{(N),II} =\,&\, -g_s f^{abc}\sum_{i+j=N-3}\eta_{ij}(\partial\overline{c}^{a})(\partial^{i}A^{b})(\partial^{j+1}c^{c})\,.
\end{align}
Note that all Lorentz indices are contracted with an arbitrary lightlike vector, i.e.,
\begin{equation}
    A^a=\Delta_\mu A^{\mu;a},\qquad \partial=\Delta_\mu\partial^\mu
\end{equation}
with $\Delta^2=0$. The operator $\Op_{c}^{(N),I}$ is required for the computation of the physical anomalous dimensions already at LO \cite{Dixon:1974ss}, while $\Op_{c}^{(N),II}$ starts contributing at NLO \cite{Dixon:1974ss,Hamberg:1991qt}. The coupling $\eta(N)$ is currently known to $O(a_s^3)$ \cite{Dixon:1974ss,Hamberg:1991qt,Gehrmann:2023ksf}. Because of the generalized (anti-)BRST symmetry mentioned above, the couplings $\eta_{ij}$ obey the following relations \cite{Hamberg:1991qt,Falcioni:2022fdm}
\begin{align}
\label{eq:conj}
    &\eta_{ij}+\sum_{s=0}^{i}(-1)^{s+j}\binom{s+j}{j}\eta_{(i-s)(j+s)} = 0\,,\\
    \label{eq:bootstrap}
    &\eta_{ij}+\eta_{ji} =\eta(N)\Bigg[\binom{i+j+1}{i}+\binom{i+j+1}{j}\Bigg]\,.
\end{align}
Note that eq.~(\ref{eq:conj}) is an example of a \textit{conjugation relation}. This can easily be checked by applying the sum in the second term to the whole expression, which leads to a mathematical identity. This type of relation was already encountered in the computation of the anomalous dimensions of the leading-twist operators in \textit{non-forward kinematics}, in which case mixing with total-derivative operators needs to be taken into account \cite{Moch:2021cdq,VanThurenhout:2023gmo}. From these studies, we know that conjugations can help restrict the \textit{function space} of the objects under consideration. To fully exploit this feature, the conjugations should be evaluated \textit{analytically}, which can be done using, e.g., the dedicated {\tt Mathematica} packages {\tt Sigma} \cite{Schneider2004,Schneider2007} and {\tt EvaluateMultiSums} \cite{Schneider:2013uan,Schneider:2013zna}.\newline

Another important consequence of the generalized (anti-)BRST symmetry is that it leads to a particular \textit{hierarchy} among the alien couplings. For example, eq.~(\ref{eq:bootstrap}) allows us to write the couplings $\eta_{ij}$, which first appear at $O(g_s)$, in terms of the leading-order coupling $\eta(N)$. In \cite{Falcioni:2024xav}, we established similar relations for the couplings of the operators that start contributing at higher orders, thus allowing us to set up a bootstrap.\newline

The power of these relations is that they allow us to severely restrict the functional form of the alien couplings. In fact, we are able to fix the couplings up to a few a priori unknown parameters, which can however be determined by computing a \textit{small} number of fixed-$N$ OMEs. This is to be compared to the hundreds of such matrix elements one would have to compute \textit{without} the relations, assuming one could somehow guess the correct function space. We have performed these computations for all one-loop alien operator couplings needed for the renormalization of the physical operators up to four loops. The explicit results can be found in Section 3 of \cite{Falcioni:2024xav}.

\section{Alien Feynman rules}
\label{sec:feynrules}
With the couplings in hand, one can derive the Feynman rules of the alien operators. In \cite{Falcioni:2024xav}, we reproduced the known results for the alien vertices with up to three external legs, which have been obtained via direct calculation in \cite{Hamberg:1991qt,Matiounine:1998ky,Blumlein:2022ndg}, as well as the four-point vertices computed in \cite{Gehrmann:2023ksf}. In addition, in \cite{Falcioni:2024xav} we also derived the five-point vertices. We find that our results are in perfect agreement with the direct calculation of the five-point vertices in \cite{Gehrmann:2024ggw}, which appeared shortly after \cite{Falcioni:2024xav}.

\section{Summary}
\label{sec:conclusion}
We have computed the all-$N$ expressions of the one-loop alien couplings which are necessary for the renormalization of the leading-twist gauge-invariant operators that define the PDFs up to four loops. This was achieved by exploiting powerful consistency relations which follow from generalized (anti-)BRST symmetry. These relations involve conjugations, which restrict the function space of the couplings. Furthermore, they showcase a hierarchy among the couplings, allowing one to express higher-order couplings in terms of simpler lower-order ones. As an application, we have determined the Feynman rules of the alien operators, and, when available, we find perfect agreement with the literature. Our results will be helpful in pushing uncertainties related to PDF evolution to the percent-level at current and future hadron colliders.

\subsection*{Acknowledgements}
This work has been supported by grant K143451 of the National Research, Development and Innovation Fund in Hungary; the EU's Marie Sklodowska-Curie 
grant 101104792, {\it QCDchallenge};
the DFG through the Research Unit FOR 2926,
{\it Next Generation pQCD for Hadron Structure: Preparing for the EIC},
project number 40824754, DFG grant MO~1801/4-2
and by the ERC Advanced Grant 101095857 {\it Conformal-EIC}.

\bibliographystyle{JHEP}
\bibliography{omebib}

\end{document}